\documentclass{article}
\usepackage[letterpaper,margin=1in]{geometry}
\usepackage{graphicx} 

\usepackage{xcolor}
\usepackage{tabularray}
\UseTblrLibrary{booktabs}

\usepackage{amsmath,amssymb}
\usepackage{color} 
\usepackage{hyperref} 
\usepackage[sort&compress,numbers]{natbib}
\usepackage{sidecap}
\title{Hi-ReX SAG summary}

\begin{document}

\begin{titlepage}
  \begin{center}
    \vspace*{1cm}
      \Huge\textbf{Science of High Resolution X-ray Imaging}
    \vspace*{0.3cm}
  \end{center}  
    
    \Large
    \noindent
    H.L.\ Marshall\textsuperscript{1}, K.\ Weaver\textsuperscript{2}, M.\ Schattenburg\textsuperscript{1}, B. Binder\textsuperscript{3} (Science Analysis Group Chairs)
    \vspace*{0.3cm}
    
    \noindent
    {\it Working Groups 1 and 2:} H.M.\ G\"unther\textsuperscript{1} (co-chair), S.J.\ Wolk\textsuperscript{4} (co-chair), K.G.\ Stassun\textsuperscript{5}, S.J.\ Gunderson\textsuperscript{1}, R.\ Pandey\textsuperscript{6,2}, L.\ Valencic\textsuperscript{6,2}
     \vspace*{0.2cm}
   
    \noindent
    {\it Working Group 3:} P.\ Draghis\textsuperscript{1} (co-chair), J.\ Hare\textsuperscript{2} (co-chair), M. Balakrishnan\textsuperscript{7}, T. Boztepe\textsuperscript{8}, 
    P.\ Gandhi\textsuperscript{9}, T.\ Holland-Ashford\textsuperscript{2}, T. Maccarone\textsuperscript{25}, H.L.\ Marshall\textsuperscript{1}, M.\ Reynolds\textsuperscript{10}, M.\ Sobolewska\textsuperscript{4}, R.\ Tanner\textsuperscript{2,11}, 
      \vspace*{0.2cm}
  
    \noindent
    {\it Working Group 4:} K.\ Weaver\textsuperscript{2} (chair), J. Cann\textsuperscript{2,12}, P.\ Chakraborty\textsuperscript{13}, B.\ Coleman\textsuperscript{2,14}, S.\
    DiKerby\textsuperscript{15},  R.\
    Gamble\textsuperscript{2,16}, J.\ Irwin\textsuperscript{17}, P.\ Maksym\textsuperscript{18}, H.L.\ Marshall\textsuperscript{1}, J.\ McKaig\textsuperscript{2,14}, E. Perlman\textsuperscript{19}, D.\ Pooley\textsuperscript{20,21}, S.
    Randall\textsuperscript{4}, H.\
    Russell\textsuperscript{22}, A.\ 
    Sarkar\textsuperscript{13}, M.\ Sobolewska\textsuperscript{4}, R.\ Tanner\textsuperscript{2,11}, S.\
    Turriziani\textsuperscript{23}, K.-W.\  Wong\textsuperscript{24}, K.\
    Whalen\textsuperscript{2,14}
   
     \vspace*{0.2cm}
   
    \noindent
    {\it Working Group 5:} S.\ Turriziani\textsuperscript{23} (co-chair), H.L.\ Marshall\textsuperscript{1} (co-chair), M.\ Balakrishnan\textsuperscript{7}, D.\ Haggard\textsuperscript{7}, T. Maccarone\textsuperscript{25}, S.\ Randall\textsuperscript{4}
    \vspace*{0.2cm}

     \normalsize   \noindent
    {\it Affiliations:} \textsuperscript{1}MIT,
    \textsuperscript{2}NASA/GSFC,
    \textsuperscript{3}Cal Poly Pomona,
    \textsuperscript{4}SAO,
    \textsuperscript{5}Vanderbilt U.,
    \textsuperscript{6}JHU,
    \textsuperscript{7}McGill U.,
    \textsuperscript{8}Istanbul U.,
    \textsuperscript{9}U.\ Southampton, UK,
    \textsuperscript{10}OSU,
    \textsuperscript{11}Catholic U.\ of America,
    \textsuperscript{12}UMBC,
    \textsuperscript{13}U.\ Arkansas,
    \textsuperscript{14}Oak Ridge Associated Universities,
    \textsuperscript{15}Michigan State U.,
    \textsuperscript{16}U.\ Md.\ College Park,
    \textsuperscript{17}U.\ Alabama, Tuscaloosa,
    \textsuperscript{18}NASA MSFC,
    \textsuperscript{19}FIT,
    \textsuperscript{20}Trinity U.,
    \textsuperscript{21}Eureka Scientific,
    \textsuperscript{22}U.\ Nottingham, UK,
    \textsuperscript{23}U.\ Antofagasta, Chile,
    \textsuperscript{24}SUNY Brockport,
    \textsuperscript{25}Texas Tech University

\vspace*{0.2cm}
 
\includegraphics[width=0.5\textwidth]{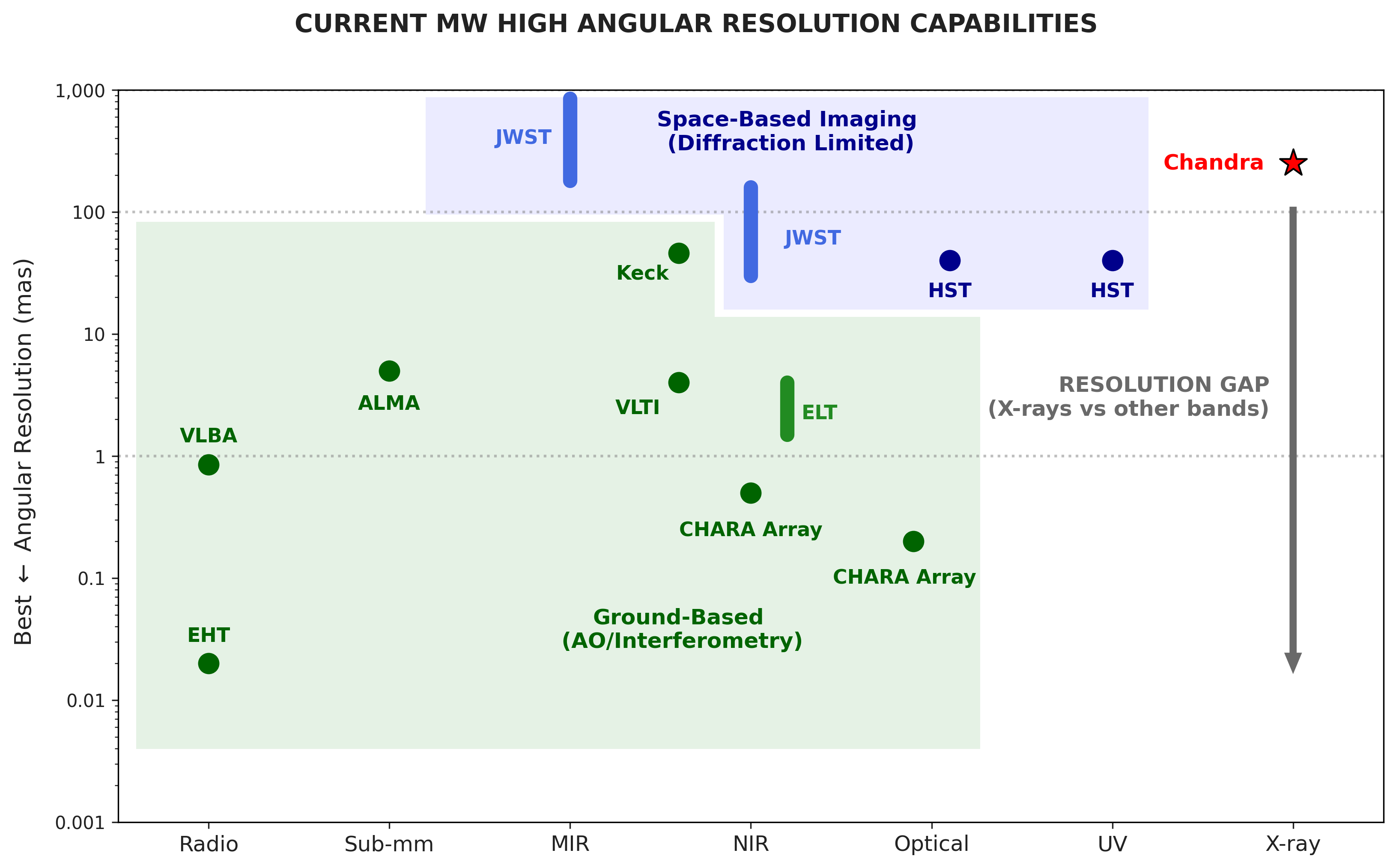}
\includegraphics[width=0.4\textwidth]{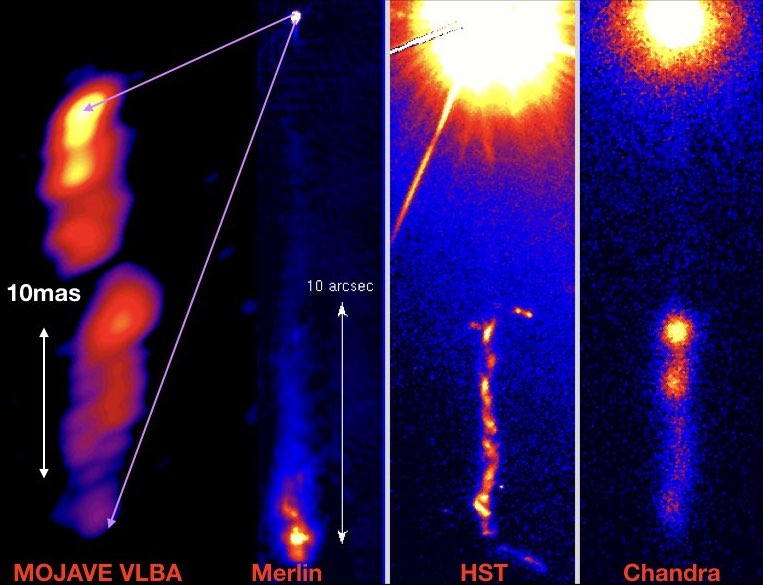}

\end{titlepage}

\begin{center}
    \Large\textbf{Science of High Resolution X-ray Imaging}
\end{center}
\vspace{-1cm}



\section{Overview}
\vspace{-0.3cm}

The aim of the Hi-ReX Science Analysis Group for an Ultra-High Angular Resolution X-ray Observatory (herein referenced as uXRI, for ultra X-ray Imager) examined many potential science cases, spanning in scale from solar system planets to high redshift quasars.  This report summarizes some of the more important potential studies.  We use ``mas'' and $\mu \text{as}$ to represent $10^{-3}$ arcsecond and $10^{-6}$ arcsecond angular scales.  An effective area of $>$500 cm$^{2}$ at 1 keV is needed for most studies.

\begin{SCfigure}[][hb]
    \centering
    \includegraphics[width=0.7\textwidth]{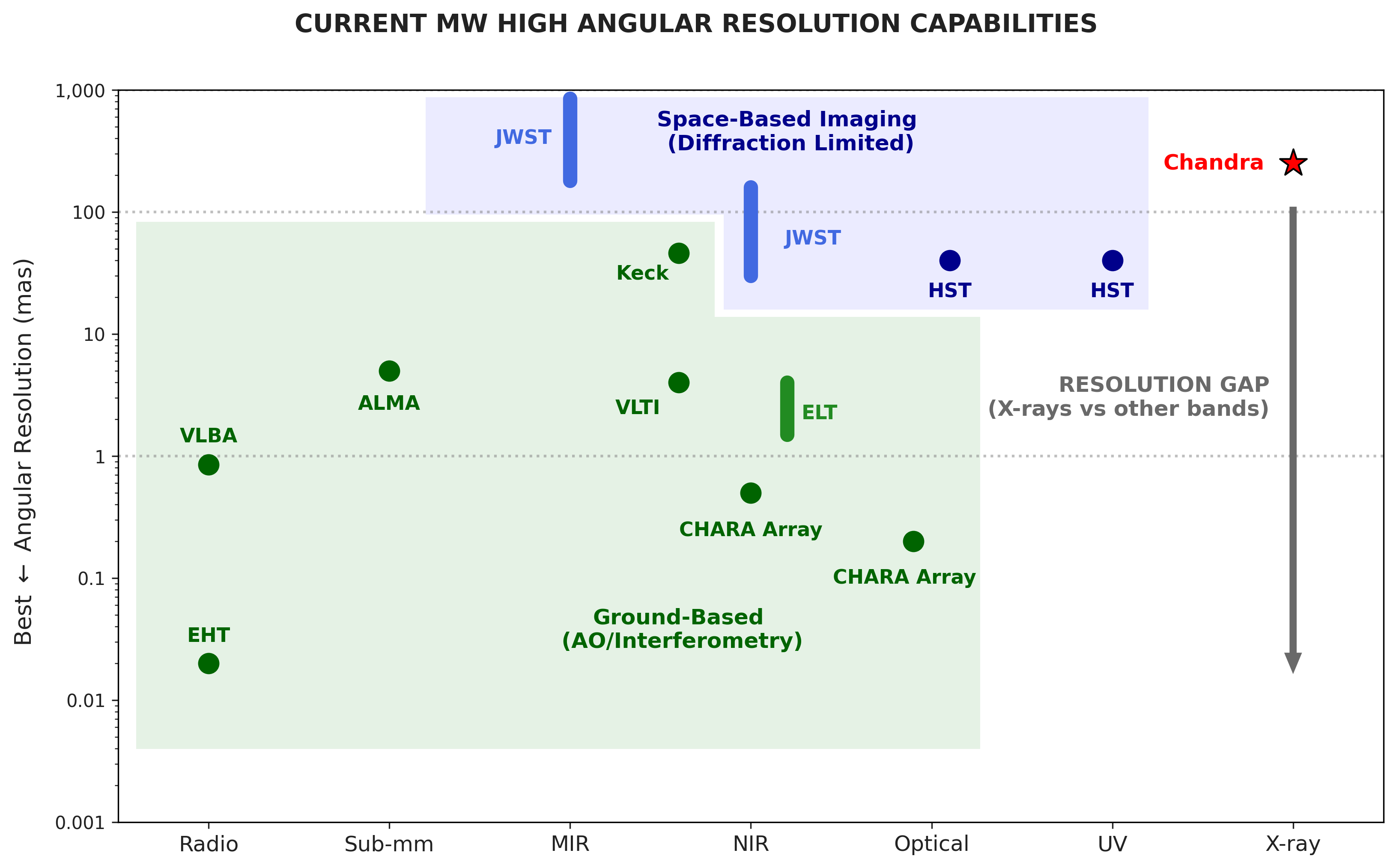}
    \caption{Comparison of the imaging capabilities of astronomical instruments in various wavebands. The best current X-ray telescope is orders of magnitude less resolution than available from the ground in the IR and radio bands. 
    }
      \label{fig:compare}
\end{SCfigure}

Fig.~\ref{fig:compare} shows the imaging gap between the current capability of the best X-ray imaging telescope, the Chandra X-ray Observatory, compared to that of instruments across the electromagnetic spectrum.  Clearly, X-ray astronomical imaging is $\times$10-10,000 from imaging at levels comparable to that in radio, IR, and optical bands.  Next-generation, ultra-high angular resolution X-ray imaging will close this gap and transform our understanding of cosmology and fundamental physics \cite{Gen-astro2010,2021ExA....51.1081U}. 

\section{Galactic Science}

\begin{figure}[ht]
    \centering
    \includegraphics[width=1.0\linewidth]{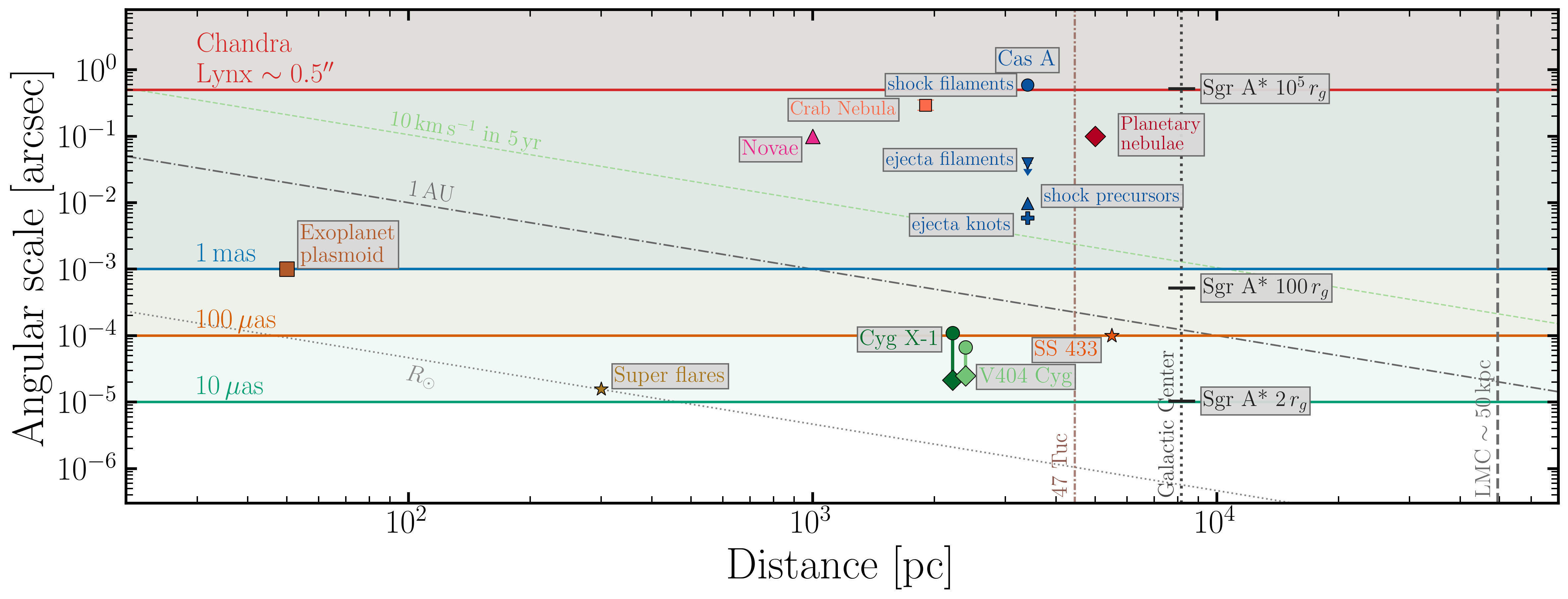}
    \caption{Estimated angular scales for science targets in the Galaxy (and its neighborhood) as a function of their distance from Earth.  Vertical dotted lines indicate the distances of targets with a variety of important spatial scales, such as the globular cluster 47 Tuc, the Galactic Center, and SN1987A.  
    }
    \label{fig:galactic}
\end{figure}

Fig.~\ref{fig:galactic} displays a broad region of the distance - angular size parameter space, focused on Galactic science. The horizontal bands show the contours and the science cases that would be unlocked by relevant milestones for the development of ultra-high angular resolution science: 1mas, 100~$\mu$as, and 10~$\mu$as. 

\begin{table}[ht]
    \centering
    \caption{Stellar Science Case Summary}
    \label{tab:starscience}

\begin{tblr}{
  colspec = {
    l
    Q[l,6.5cm]
    l
    l},  
  row{4,6, 10} = {gray!20},
  row{1}    = {font=\bfseries, bg=white},
  cells = {valign=m},
}
\toprule
Topic & Science Case & Resolution & Other Requirements \\
\midrule
        Solar System & Resolve comets & 100 mas & large $A_\mathrm{eff}$\\
        & Planetary Aurorae & 5 mas & \\
        Planetary Nebulae & Observe mixing in fast wind and nebula, radial distribution of hot bubble mission & 100 mas & 0.1-1 keV, arcmin field\\
        Star Formation & Proto-stellar jet launching, distant stellar clusters & 10 mas & 0.5$-$2.0 keV \\
         (Exo-)planets   & Exoplanet plasmoid & 1 mas & \\
        Activity & Astrospheres & 100 mas & 0.2$-$2.0 keV \\
            & Coronal superflares & 10 $\mu$mas & arcminute field\\
            & Extreme coronal mass ejections (CMEs) & 10 $\mu$as & \\
        Massive Stars     & Colliding wind binary shock cones & 100 $\mu$as & \\
    \bottomrule
     \end{tblr}   
\end{table}

\subsection{Pathways to habitable worlds}
Exoplanets are obviously the places where we look for life in the universe and it is their host stars that can make planets hospitable to life or destroy all hope of finding it. The formation of stars and planets is intricately linked with planets evolving in the disks that surround young stars. High-resolution X-ray imaging can resolve hot, ionized regions in these systems, such as host-star coronae, and determine how they move and how they impact the evolution of stars, planets, and possible life.

\subsubsection{Solar and stellar winds}
The only planetary system that we know for sure to be habitable is our own. It is thus critically important to study how energetic radiation interacts within our system to understand what impact it has on planets, atmospheres, and biomolecules in space.
Planets and comets in our own solar system emit X-rays via charge exchange with the solar wind. A uXRI mission can remotely resolve features that are currently limited to very rare and expensive in-situ measurements (e.g., Juno around Jupiter) to study space weather, atmospheres, and mass loss. The brightest of those phenomena are also observable in exoplanet systems with sufficient spatial resolution, where they open a window into processes that will never be available to in-situ studies.

The region filled by a host star's wind is called its astrosphere. It can be detected in the 0.1-1 keV range \citep{2024NatAs...8..596K,2026ApJ...999..125L}
as the highly ionized stellar winds interact with neutral gas \citep{2019ApJ...886...41L}.
Our Sun is obviously the best-studied system \citep{2012AN....333..324D}.
The winds of Sun-like main-sequence stars are relatively modest but have a tremendous impact on carving out a host-star-dominated bubble in the galactic ISM. Inside this bubble, surfaces and atmospheres of bodies are directly affected by the stellar wind,
not the ISM. Stellar winds can erode planetary atmospheres through sputtering and ion pick-up, or redden, reduce, and erode airless surfaces through ``space weathering''.
At a resolution of 100 mas we could resolve solar-like astrospheres out to 100 pc and determine their shape, which is an indication of wind strength.  There are over 150,000 stars within this volume that can be examined, of which about 1000 are visible to the naked eye and $>$ 10,000 are of type G. 

\subsubsection{Cool stars and their planets}
\begin{SCfigure}[][ht]
    \centering
    \includegraphics[width=0.4\linewidth]{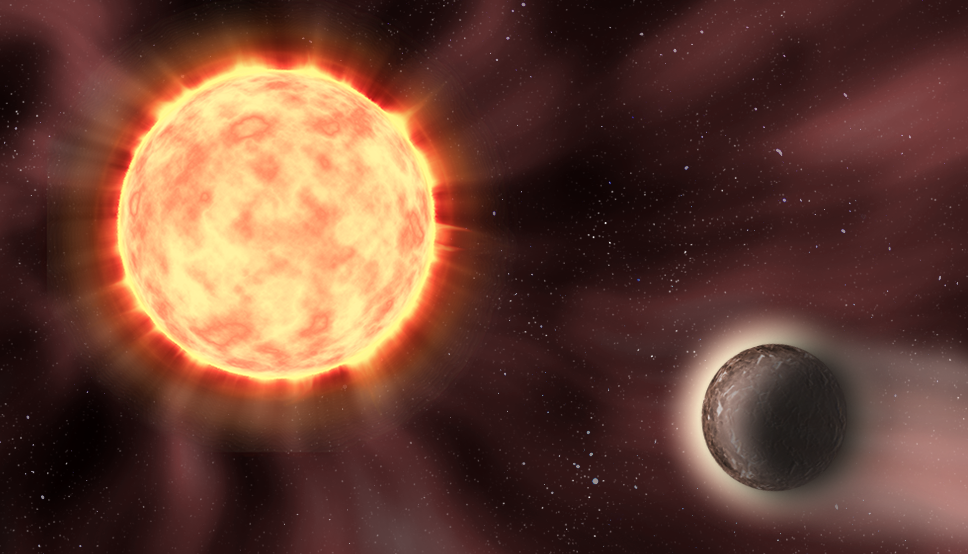}
    \caption{Artist's illustration of a star system where the stellar wind
    is ablating its planet's atmosphere.  The X-rays produced at the stellar surface can be resolved from that produced at the planetary atmosphere with a uXRI with 10 $\mu$as imaging resolution.  Image is from a press release \cite{2023MNRAS.524.5060C}.}
    \label{fig:planetablation}
\end{SCfigure}

X-rays from cool stars are often the best probes of their magnetic fields.
For 
stars like our Sun, the magnetic field
can power coronal mass ejections (CMEs) and X-ray flares.
Stellar X-rays can change the chemistry in their
planets \citep{2022A&A...667A..15K}, evaporate the atmospheres of close-in planets, as imagined in Fig.~\ref{fig:planetablation} \citep{2018MNRAS.479.5012O,2023MNRAS.524.5060C}, and destroy complex molecules and prevent the evolution of life \citep{2019ApJ...881..114Y}.
Long-duration X-ray flares in pre-main-sequence stars have magnetic fields connecting the stellar surface to the inner disk, affecting planet formation \cite{2005ApJS..160..469F,2008ApJ...688..437G,Aarnio2012}.
To resolve superflares and CMEs and pinpoint them to the star, we need to resolve the star from the inner disk
requiring an imaging resolution of $\sim 10\;\mu$arcsec.

\subsection{Distributing the Elements of Life}

\subsubsection{Planetary Nebulae}

Planetary nebulae (PNe) mark the transition of low-to-intermediate mass stars (1-8 M$_{\odot}$) from the asymptotic giant branch to white dwarfs. During this phase, the star donates at least half of its metal-enriched total mass to the ISM. 
Diffuse X-ray emission in PNe is understood to originate in the hot bubble formed by the fast winds of the central star
\citep[e.g.][]{2016ApJS..226...15F}.
The radial distribution of hot bubble emission \citep{2016ApJS..226...15F} and mixing of nebular and fast wind abundances \citep{2003PASP..115.1002M} cannot be seen with current X-ray facilities. A uXRI with spatial resolution $\le$ 100 mas will resolve the radial distribution of hot bubble emission and detect differences in elemental abundances essential to life.

\subsubsection{Massive Stars}
Massive stars, i.e., spectral types OB and Wolf-Rayet (WR), are intrinsic X-ray sources due to hypersonic shocks in their winds with speeds of $v_\infty=1500 - 3000$\,km\,s$^{-1}$,
altering both the surrounding interstellar medium (ISM) and galactic feedback \citep{Krause2013}.
There are three broad categories for the types of wind shocks: embedded wind shocks (EWS), colliding wind shocks (CWS), or magnetically-confined wind shocks (MCWS).
The X-ray emission from the EWS constrain important parameters like the mass-loss rate 
but the distribution of shocks in the wind is not known \cite{Owocki2004}. A uXRI with spatial resolution $\leq0.1$\,mas would resolve the shock distribution and test the wind acceleration and clumping sources, crucial for determining the mass-loss rates and understanding their contribution to ISM chemical enrichment.
CWS  
form between the extended winds of two stars \citep{Rauw2016}. In the most dramatic case, the surface is referred to as a ``shock cone'' that wraps around the star with a weaker wind. An imager with spatial resolution $\leq0.1$\,mas would resolve not only the shock surface but also the individual EWS of the binary components.
Like in PNe, the mass loss rates set the amount of turbulence that sources cause in the ISM, how they disrupt star formation, and shape their galaxies.

\subsection{How are particles accelerated?}

\subsubsection{Supernova remnants SNRs}

SNRs are thought to be responsible for cosmic ray acceleration up to the ``knee'' at $\sim$3 PeV.
However, it is currently unclear whether the rapid dropoff in downstream synchrotron emission is due to electron radiative losses or magnetic damping \citep{2014ApJ...790...85R,2015ApJ...812..101T}.
While filament widths have been measured down to $\sim 3 \times 10^{16}$ cm
\cite{2005ApJ...621..793B}, the theoretical scale length for an electron radiating at 4~keV can be as small as $5 \times 10^{14}$ cm: 10 mas at D=3.4~kpc. To study the full structures and energy-dependent widths of shocks in Galactic SNRs up to 8~kpc away, a uXRI with $\lesssim$10 mas is necessary (see the blue triangle data point at D=3.4~kpc in Fig.~\ref{fig:galactic}).

\subsubsection{Relativistic jets and outflows}
\label{sec:xrbjets}

The highest energy cosmic rays are often thought to originate within relativistic jets in X-ray binaries involving neutron stars or black holes. 
SS 433 is the clearest case in the Galaxy, producing TeV photons \cite{2018Natur.562...82A} from acceleration sites with high X-ray polarization \cite{2024ApJ...961L..12K} and showing knots of radio emission with proper motions of about 7 mas/day \cite{2013ApJ...775...75M}.
X-ray spectroscopy of SS 433 shows emission lines of common elements associated with its twin jets, due to entrainment of surrounding material or interaction between initially leptonic jets and dense ambient gas.  Given a cooling time of $\sim 5000$ s \cite{1996PASJ...48..619K,2013ApJ...775...75M,2026PASJ...78..436S} the jets will be resolved by a uXRI with resolution of 0.1 mas.
Higher resolution images of the hotspots in its nebula should isolate the regions that produce TeV photons, much as VLBI images of the Pictor A hotspot shows features on mas scales \cite{2008AJ....136.2473T}.
More generally, resolving transient X-ray jet ejecta close to launch would allow kinetic measurements of the outflow, direct constraints on jet morphology and deceleration, and spatially resolved spectroscopy of the emitting plasma, as carried out with Chandra and VLBI at scales from 1-10,000 mas \cite[e.g.][]{2002Sci...298..196C,2024ApJ...971L...9W}.
\subsection{How do stars die? }

When compact objects are formed, they may exit their systems due to asymmetries of the supernovae that produced them. Neutron stars typically have large velocities, of order $400\;{\rm km/s}$, constraining supernova explosion physics \citep{2005MNRAS.360..974H}.
Current X-ray proper-motion measurements require are limited to small, nearby, bright sources. Milliarcsecond X-ray astrometry would greatly expand this sample: $\sim 1$ mas localization would measure Galactic neutron-star velocities to $\sim 50\;{\rm km/s}$ over year-long timelines. Such measurements would test whether neutron stars move opposite to bulk ejecta, whether kick magnitudes correlate with SNR properties, and how often extreme $>1000\;{\rm km/s}$ kicks occur.
The birth sites of magnetars (pulsars with the highest magnetic fields) are not well determined, so the progenitor type is unknown -- a problem solvable with uXRI observations of proper motions at the mas/year level \cite{2026arXiv260315750C}.
The same capability is crucial for black-hole X-ray binaries, where natal kicks remain debated: some systems require strong kicks, while others may be consistent with weak recoil from fallback \citep{zhao23}. High-resolution X-ray imaging would discover and track compact remnants in Galactic regions too obscured for optical observations, providing the astrometric sample needed to determine how compact-object kicks depend on remnant mass and explosion physics.

\subsection{How is accretion fueled? } 

\subsubsection{Feeding from the ISM}
\label{sec:sgra_bondi}

Sgr A* is the nearest laboratory for understanding radiatively inefficient accretion onto a supermassive black hole (SMBH).  Chandra can resolve the hot gas reservoir near the Bondi radius \cite{Bon52} at $\sim 4''$, which is where the gravity of the SMBH dominates gas motion.
However, the crucial region connecting the gas supply to the inner accretion flow remains unresolved. The key problem is that stellar winds supply gas at roughly
1000 times the rate that reaches the SMBH \cite{2008MNRAS.383..458C,2020ApJ...896L...6R}.
A mas uXRI would resolve scales where simulations predict a transition from quasi-spherical inflow to a rotating, outflow-dominated structure and measure the density, temperature, and velocity structure of the flow and determine how Sgr A* starves despite the available gas supply. 

\subsubsection{Feeding from a star's surface}

In many X-ray binaries, matter flows directly from the surface of a companion star to the compact object through a point where the gravitational pull of the two stars is equal, called Roche lobe overflow.
The physical scale of the transfer is comparable to the size of the binary, resolvable at $10-100\;\mu$as with a uXRI.
X-ray imaging would begin to resolve the physical scale on which mass is transferred into accretion disks in compact binaries.
Using V404 Cyg as a fiducial system ($d=2.39\;{\rm kpc},\;M_{\rm BH}=12\;M_\odot$; \cite{1994MNRAS.271L..10S, 2009ApJ...706L.230M, 2010ApJ...716.1105K}), the binary separation is $\sim0.16$ AU, corresponding to an angular scale of order $100\;\mu$as (green circle in Fig.~\ref{fig:galactic}).
Such observations would constrain how mass transfer feeds the disk and how these systems accumulate enough material to go into outburst.

\subsubsection{Feeding from a star's wind}

In high-mass X-ray binaries mass is captured from the massive companion’s stellar wind. Cygnus X-1 \cite{2021Sci...371.1046M} is the key fiducial case: the binary separation is about 0.4 AU, so
a $10\;\mu$as uXRI would resolve scales of $0.02$ AU at 2.2 kpc.
(The green circle and diamond in Fig.~\ref{fig:galactic} show the angular size of the binary separation and of the accretion disk in Cygnus X-1, respectively.)
Such an observation would spatially connect the black hole environment to the companion's wind showing how the asymmetric stellar wind converts into an accretion flow, whether the inflow forms a stable outer disk or remains wind-fed and variable, and how changes in the donor wind propagate inward to modulate the X-ray luminosity.


\subsection{What are the populations of X-ray sources?}

\subsubsection{At the Galactic Center}

Deep Chandra studies have shown that the central few degrees contain thousands of discrete X-ray sources, while work on the Galactic ridge and bulge suggests that emission once treated as diffuse can largely be decomposed into faint point sources, plausibly dominated by magnetic cataclysmic variables, accreting white dwarfs, active binaries, and ultracompact X-ray binaries.
The average angular separation of stars at the Galactic center is about 8 mas, so imaging with uXRI at 1-2 mas would identify each X-ray source with an IR counterpart, directly testing the suggestion that the Galactic Center X-ray emission is dominated by intermediate polars and related accreting white-dwarf systems \cite{2009Natur.458.1142R,2016ApJ...826..160H,2009ApJS..181..110M,2003ApJ...589..225M}.

\noindent
\subsubsection{Astrometry in globular clusters}
Globular clusters host an overabundance of X-ray binaries compared to the Galactic disk, however, it is often difficult to identify the true optical/IR counterpart to the X-ray source due to positional uncertainties of $\approx0.3''$ or larger \cite[cf.][]{2018ApJ...865...33H,2018MNRAS.479.2834H}. A X-ray observatory with an angular resolution of 10 mas or better would measure the X-ray source positions
well enough to obtain
a confident determination of the true optical/IR counterpart (or lack thereof) to the X-ray source. Identifying the true counterparts of the X-ray sources is critical for source classification, particularly at the faint end of the X-ray luminosity function. Additionally, an angular resolution of $<1$ mas would allow for the measurement of X-ray proper motions. Several prominent clusters (e.g., 47 Tuc, Omega Cen, Terzan 5; \cite{2018A&A...612A.115N,2018ApJ...854...45L,2015ApJ...810...69M}; see Fig.~\ref{fig:galactic}) have proper motions of several mas yr$^{-1}$, which would easily be measurable and even the internal motion of X-ray sources (with typical velocities of $\sim 10-20$ km s$^{-1}$; \cite{2019ApJ...875....1M,2019MNRAS.482.5138B}) could be measured across a $\sim5$ year baseline. This would allow for one to determine cluster membership based solely on the X-ray proper motion, enabling a complete census of X-ray source populations and a study of their kinematics,
necessary for direct comparison to simulations (see e.g., \cite{2026enap....3..458K} for a recent review). Searches for fast moving X-ray sources could also be carried out, similar to those recently discovered in HST observations of Omega Cen \citep{2024Natur.631..285H}, which were used to place constraints on the location and mass of a potential intermediate mass black hole in the cluster.

\section{Extragalactic Science}
Key science cases for high-resolution X-ray imaging for extragalactic celestial sources
are given in Table \ref{tab:science_summary}.


\begin{table}[ht]

\centering
\caption{Extragalactic and AGN Science Case Summary}
\label{tab:science_summary}

\begin{tblr}{
  colspec = {
    l
    Q[l,6.5cm]
    l},  
  row{2-3, 7-10, 14, 16} = {gray!20},
  row{4-6} = {white},
  row{1}    = {font=\bfseries, bg=white},
  cells = {valign=m},
}
\toprule
Science Theme & Goal & Resolution \\
\midrule
AGN jets & Launching mechanism & $\mu \text{as}$ \\
         & Acceleration sites & mas \\
AGN outflows & {AGN feedback in\\ protocluster evolution} & 10-30 mas \\
& {Feedback-regulated cooling and star formation} & $\mu$as\\
& {Accretion Disk Winds} & $\mu$as \\
AGN Accretion and Corona & Corona morphology & $\mu$as \\
 & {Constrain binary SMBH merger rates} & 0.1-10 mas \\
& {Final pc problem} & 120 $\mu$as \\ 
& Characterize accretion flow at scales $<0.1 R_{\rm B}$ & $\sim10\mu$as \\
General Relativity  & {Orbiting hotspots in Sgr A*}   & $10 \mu \text{as}$ \\
        & Disk imaging in Fe-K$\alpha$   &  $0.1 \mu \text{as}$ \\
        & BH event horizon   &  $0.1 \mu \text{as}$ \\
Dark matter   & Proper motions of galaxies  & mas \\
Cosmology & {Intracluster medium microphysics}    & $50 - 200 \text{mas}$ \\
Gravitational Waves & {Identify sources of nHz waves}    & $\mu \text{as}$ \\
\bottomrule
\end{tblr}
\end{table}

\begin{figure}[ht]
    \centering
    \includegraphics[width=1.0\textwidth]{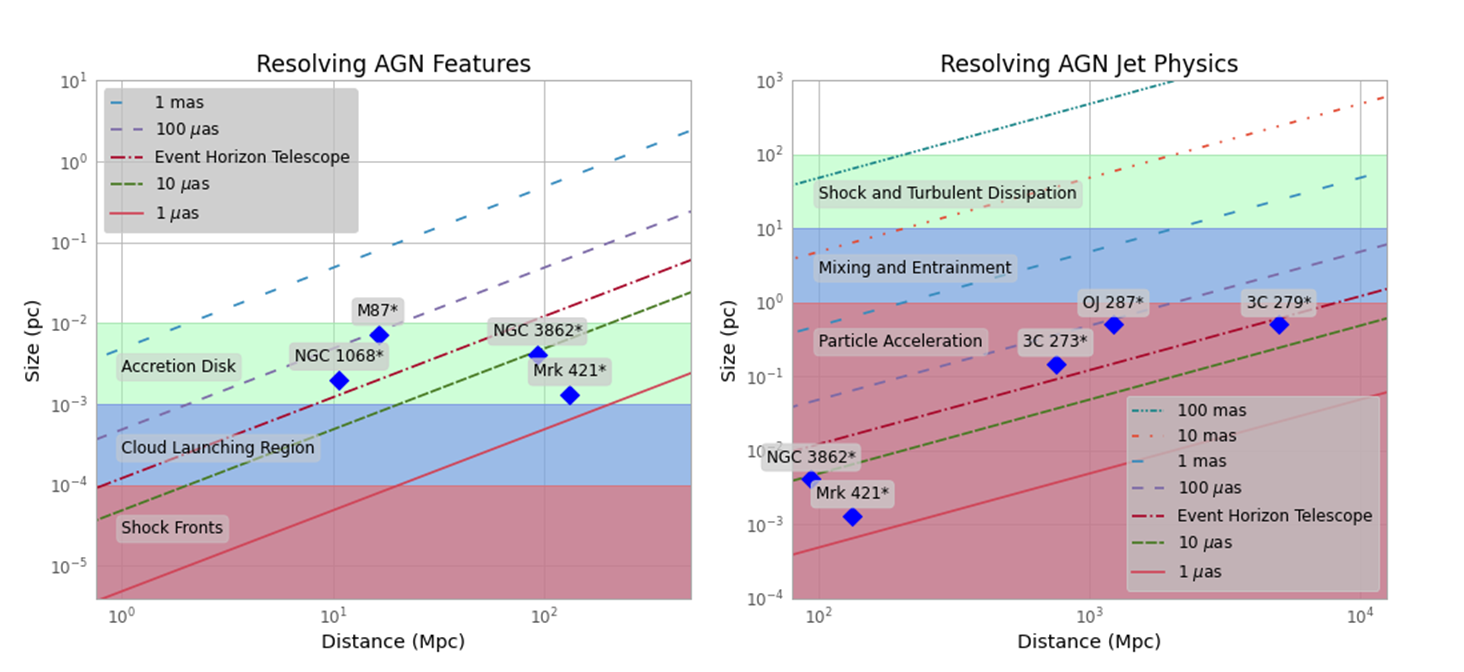}
    \caption{Left: Angular scales needed to resolve the accretion disk systems of nearby AGN, marked with upper limits. Right: Angular scales (diagonal lines) needed to address AGN jet physics at various distances. Lower limits to the accretion disk/jet launching region sizes are noted for various AGN.
    }
      \label{fig:resolution}

\end{figure}

\subsection{Active Galactic Nuclei (AGN) Accretion and Coronae}
\subsubsection{AGN corona size and geometry}
\label{sec:agncorona}

For the dominant population of radio-quiet AGN, the X-ray emission is not from the accretion onto the SMBH (which emits predominantly in the optical or UV bands) but from a region above the disk, the so-called ``corona'', comprising a substantial fraction of the total electromagnetic power.  X-rays from the corona reflect off of the disk, producing a relativistically broadened Fe line that has been used to estimate the spins of AGN SMBHs \cite[cf.][and references therein]{2021ARA&A..59..117R}.
While X-ray reverberation studies have pointed to a compact corona above the disk located within tens of gravitational radii of the SMBH \cite[cf.][and references therein]{2021iSci...24j2557C}).  Polarimetric results indicate a slab- or wedge-like corona in one case with a solid detection \cite{gianolli2023,gianolli2024}. Spectroscopic analyses remain extremely model dependent (e.g., \cite{Tatum2013, Marinucci2014}). A uXRI with $\mu$as resolution will obtain direct images of the coronae in nearby AGN, showing their true geometry and relationship to the accretion disk, as well as providing kinematic results for testing models of the local spacetime (see Fig.~\ref{fig:resolution} left and \S~\ref{sec:gr}).
This study can be extended to high $z$ sources 
with the use of gravitational lensing \cite{1986ApJ...301..503P}.

\vspace{-0.2cm}
\subsubsection{Sub-Bondi radius accretion \& gas flow}
\vspace{-0.1cm}

As with Sgr A* (\S~\ref{sec:sgra_bondi}),
the Bondi radius characterizes the transition from the ambient interstellar medium to the start of material infalling into the SMBH. 
The Bondi radius has only been resolved in X-rays for the five closest \cite{2024Univ...10..278W}, all low-luminosity AGN (LLAGN) believed to be accreting in the hot accretion mode. However, significant populations of LLAGN have been discovered at high redshift (e.g., \cite{Lin2026, Geris2026}) and the properties of their Bondi accretion flow remain unknown. With a resolution of $\sim 10 \mu$as, a uXRI would be used
to test whether the temperature structure is anisotropic, as predicted by some inflow-outflow simulations, or more random, as expected in chaotic accretion models, providing key insight into the accretion processes powering AGN in their earliest epochs and across cosmic time. 

\subsection{AGN Outflows and Feedback}


\subsubsection{The Effect of AGN on Galaxy Evolution}

AGN feedback is important in regulating star formation, cooling, and outflow rates in galaxies, but the details of how  AGN activity affects the ISM are not straightforward. 
Momentum-driven and energy-driven outflows have significantly different implications for the transfer of energy from the outflow into the surrounding ISM \citep[see][for a review]{2015ARA&A..53..115K}.
Momentum-driven shocks are inefficient, transferring only $<$1\% of the AGN luminosity to the ISM, compared to the much more efficient energy-driven regime where the efficiency is $\sim15\%$. 
Outflows in the region immediately surrounding the AGN will be momentum-driven, 
before ultimately transitioning into energy-driven outflows at further distances. This transition region is highly dependent on the type of outflow (wind vs. jet), the ISM, and the rate at which the individual shock layers will break down due to turbulent mixing, magnetic fields, and cooling rates.
A uXRI with mas to $\mu$as resolution could resolve the cloud launching region and the various layers of the shock fronts around the AGN bubble and around the cool, dense clouds embedded in the wind.  These clouds may be responsible for the ubiquitous ultrafast outflows reaching $0.3-4 c$ \cite{2024A&A...687A.179X}.


\subsubsection{The Effect of AGN on Galaxy Cluster Evolution}

The origin of the thermodynamic and chemical structure of galaxy clusters remain one of the central unresolved problems in cluster astrophysics. In the standard picture, the intra-cluster medium (ICM) is shaped over cosmic time by AGN heating, gas uplift, shock driving, turbulence, and metal transport \cite[e.g.,][]{2007ARA&A..45..117M,2012ARA&A..50..455F}.
The remarkably uniform metal abundances observed in clusters and groups
strongly suggest that much of this enrichment and mixing may have occurred early, at $z\sim 1-3$ \cite[e.g.,][]{2017MNRAS.470.4583U,2022MNRAS.516.3068S}, when protoclusters were still assembling and powerful AGN outflows distributed energy and metal-enriched gas throughout the nascent ICM. 
A uXRI with $10-30$~mas resolution can directly image X-ray shocks, cavities, outflow cones, and metal-enriched clumps in high-redshift proto-cluster environments on scales of 100s of pc. These structures are the fossil record of how AGN mechanical energy is thermalized. By mapping small-scale metal-rich plumes, such observations will directly constrain metal transport by AGN gas uplift, galaxy-scale winds, ram-pressure stripping, or turbulent mixing. This will also connect early SMBH growth to the later entropy floors, metallicity profiles, and scaling relations. 

\subsubsection{Jet Launching and Collimation}

\begin{SCfigure}[][ht]
    \centering
    \includegraphics[width=0.55\textwidth]{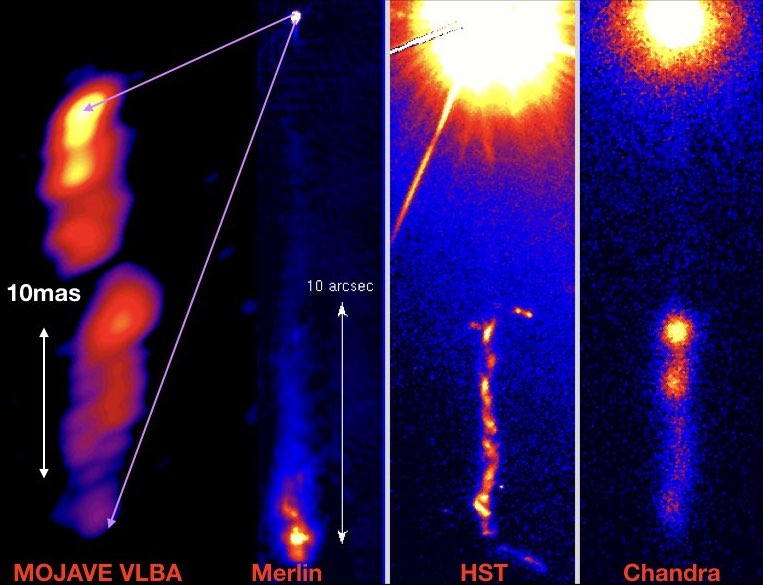}
    \caption{ Images of 3C 273 in several bands, where 10 mas corresponds to about 34 pc.  Left: radio band images from a snapshot of a VLBA movie taken from the MOJAVE web site (\text{https://www.cv.nrao.edu/MOJAVE}). Right three images: radio, optical and X-ray images \cite{2001ApJ...549L.167M}. Note that there is clearly structure in the optical band at 100 mas resolution that is not observable with Chandra.  Also, the core is well resolved into a pc-scale jet at the few mas scale.  A uXRI with 5 mas imaging would resolve both the core and the acceleration sites in the kpc scale jet. }
    \label{fig:3c273}
\end{SCfigure}

As in X-ray binaries (\S~\ref{sec:xrbjets}), AGN jets play an important role in the accretion process, where they dissipate angular momentum and allow matter to fall closer to the black hole, while also transporting energy to scales of kiloparsecs, where they affect the larger galaxy and cluster environments.
The acceleration, collimation and stability of jets are still open questions in modern astrophysics \citep{Blandford2019}. The internal jet structure at scales of tens to hundreds of parsecs is key to identifying the macroscopic characteristics of kinetically dominated jets that impact their surrounding environment \citep[e.g., M87;][]{Lobanov2003}.
While studies using Chandra have resolved many kpc-scale jets \cite{2018ApJ...856...66M} such as 3C 273 (Fig.~\ref{fig:3c273}), there are indications from variability studies that these jets have substructure on scales at 1-100 mas or smaller \cite{2010ApJ...714L.213M,2023NatAs...7..967M}, resolvable with a uXRI with mas-scale resolution to examine particle acceleration sites. Parsec-scale jets are already known to be quite dynamic at 1 mas scales, as observed with VLBI but X-ray emission arises from electrons with much shorter lifetimes, so their emission occurs much closer to acceleration sites (Fig.~\ref{fig:3c273}).
A uXRI with $\sim 5-10$ $\mu \text{as}$ resolution would allow us to study the smallest scales on which material has been observed to flow into the jet, providing key context into the dynamic processes behind jet launching and collimation (Fig.~\ref{fig:resolution} right). 
Moreover, if this mission is complemented by polarimetric measurements, it will permit us to map the magnetic field along the jets, investigating the presence of transient electric fields, which can accelerate charged particles in the jet to the highest energies \citep[see e.g.][and references therein]{SKA_chap2015-ST}. The Blandford-Znajek process requires that the jet magnetic field winds about on scales comparable that of the event horizon, which can only be determined with X-ray polarimetry.

\vspace{-0.2cm}
\subsection{Dual and binary AGN}
\vspace{-0.1cm}

The merger scenario for galaxy and SMBH formation implies that there should be galaxies with several SMBHs \citep{WR1978}.  Historically, they have been referred to as (i) dual systems with black holes that are gravitationally unbound
or (ii) binary systems that are gravitationally bound. 
Even triple systems have found \cite{Djorgovski2007,Schwartzman2024}.
The frequency of multiple SMBH systems has implications for merger rates and thus for black hole growth history. 
Close binary SMBHs are sources of gravitational waves, relevant to the stochastic gravitational wave background \citep{RR1995}, NANOGrav \citep{NanoGrav15-2023}, and LISA \citep{LISA2017}.
A uXRI with $\sim10$mas to 100\,$\mu$as resolution will allow us to confirm binary (and dual/multiple) AGN candidates selected at other wavelengths down to parsec scale separations at $z \leq 1$.


\subsection{Testing Theories of Gravity}
\label{sec:gr}

Since X-rays are generated by matter under extreme physical conditions, such as the vicinity of compact objects or in the centers of AGN, they allow us to study general relativity (GR) in the strong-field regime by analyzing the emission arising in the extreme gravitation field of black holes. A uXRI mission holds the potential to open a new window in the testing of GR and alternative gravity theories near black holes, offering direct evidence of spacetime warping and accretion inflows and outflows.
Testing GR with electromagnetic tests is complementary to the studies performed with gravitational waves (GWs).
GWs probe the ``dynamic'' spacetime in binary systems during the coalescence phase.
By contrast, electromagnetic tests are sensitive to the motion of particles
in ``static'' spacetime, addressing phenomena such as the bending of light and gravitational redshift \citep{Bambi2017, 2022hxga.book...81A}.

\subsubsection{Imaging Orbital Motion}

\begin{SCfigure}[][ht]
    \centering
    \includegraphics[width=0.6\textwidth]{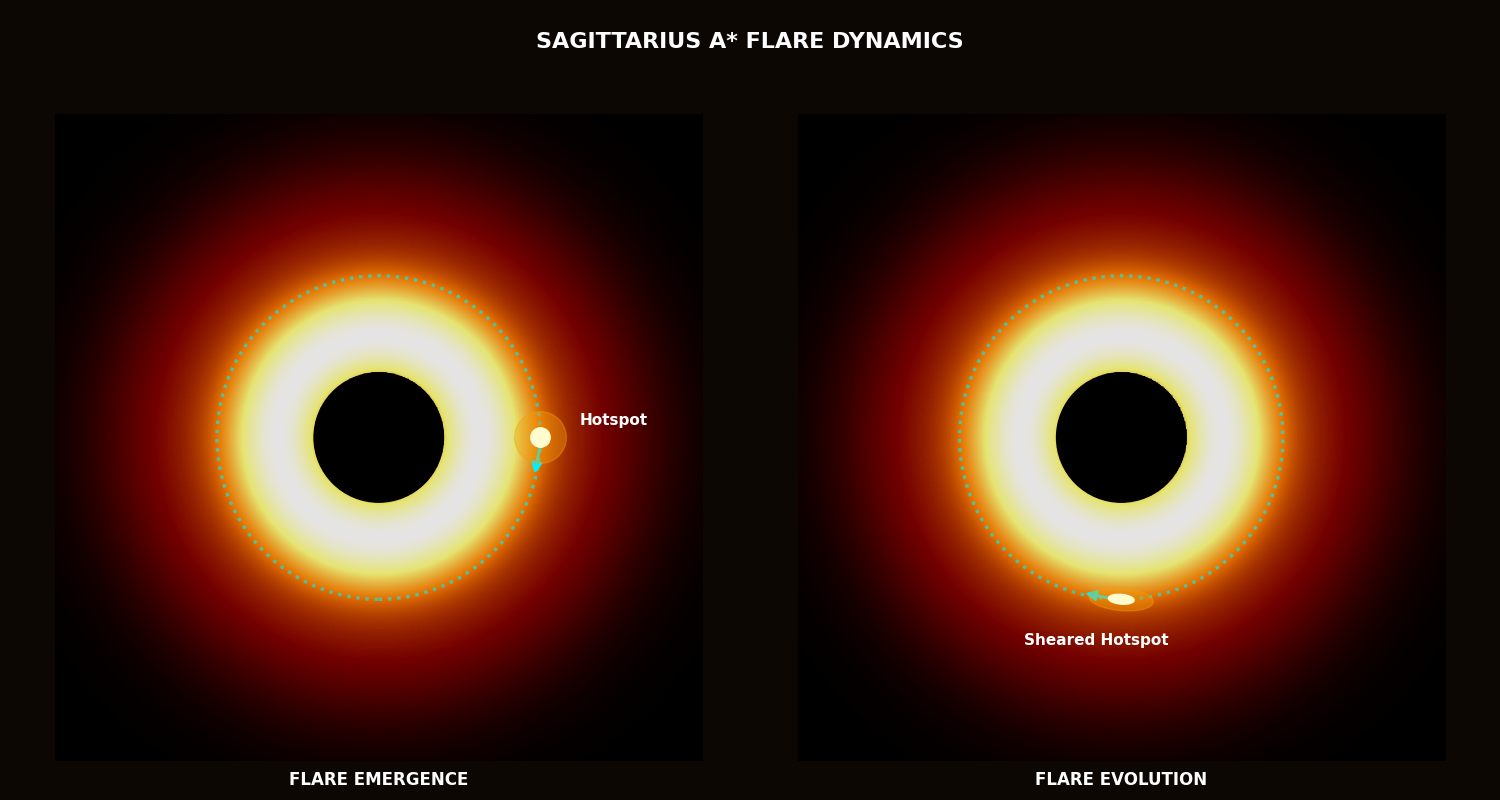}
    \caption{ {Schematic representation of an orbiting hotspot near Sgr A*. \textbf{Left panel ($T=0$)}: An idealized, spherical hotspot emerges in the inner accretion disk in connection of an X-ray flare.  \textbf{Right panel}: The hotspot advances along its orbit (by a $90^\circ$ clockwise rotation), with elongation and  morphological distortion induced by intense gravitational shear near the event horizon.}}
    \label{fig:hotspot}
\end{SCfigure}

\noindent
Sgr A* is practically a dormant black hole, accreting at a very low rate through a radiatively inefficient accretion flow.  While accretion onto the BH in Sgr A* can be studied at the scale of the Bondi radius (\S~\ref{sec:sgra_bondi}), some X-ray emission occurs much closer to the event horizon, evidenced by
X-ray flares \citep[e.g.][]{Dodds-Eden2011,Witzel2021}.
Current modeling indicates that they can be explained with synchrotron or synchrotron-self Compton emission from non-thermal electrons \citep{Dodds-Eden2009,Ponti2017,GRAVITY2021}. 
A $10\mu\text{as}$ uXRI can detect flaring hotspots orbiting Sgr A* near the innermost stable orbit (ISCO) about the black hole, allowing measurements of the orbital trajectories in the inner part of the disc to be compared with the predictions of general relativity and alternative theories of gravity \citep{Shahz2022}.

\subsubsection{Imaging the Accretion Disk}

While most of the X-ray light is from the disk's corona (\S~\ref{sec:agncorona}), the disk is visible in X-rays by way of reflection.
The relativistically broadened component of the Fe K$\alpha$ line is a reflection feature that provides information on the dynamics, thermodynamics, and orientation of the disk very close to the SMBH event horizon.
The presence of the SMBH introduces relativistic effects that are strongest close to the black hole.  The combination of relativistic Doppler effect and gravitational redshift results in a highly broadened and skewed line profile that is sensitive 
to spin and inclination angle of the disc, which depend on the local spacetime.
A uXRI with 0.1 $\mu$as resolution at 6 keV would resolve the motion of Fe-K$\alpha$ emission as the gas orbits the SMBH.

\subsubsection{Imaging the Supermassive Black Hole Event Horizon}

The Event Horizon Telescope (EHT) produced classic images of the event horizon ``shadows'' around Sgr A* and M87.
These two targets were ideal for the EHT due to SMBH mass and radio intensity.  They are not, however, very bright X-ray sources and are also embedded in X-ray bright hot diffuse gas.
Still, there are AGN for which a uXRI can be effective.
Estimating the ISCO sizes for AGN in the ETHER sample \citep{ETHER2023} and AGN with reverberation mapping mass estimates, there are some radio quiet AGN (that are not jet dominated) have apparent sizes of order 0.2-0.1 $\mu\text{as}$, requiring a uXRI with a resolution of better than 0.1 $\mu\text{as}$.

\subsection{Probing dark matter distributions}
Understanding the dark matter properties of nearby galaxies and clusters would be vastly improved with full phase space information.
Assuming accurate true distances, proper motion measurements would yield transverse velocities -- the last of six phase space components.
Proper motions of AGN in the Bullet Cluster, for example, can reveal the relative masses of the two colliding clusters \cite{2018ASPC..517..663M}, and relative proper motions of galaxies in groups can reveal whether the dynamical friction expected from dark matter models, but not expected for Modified Newtonian Dynamics models, is taking place \cite{2017MNRAS.467..273O}.
Proper motions with accuracies of about 1 $\mu$sec/year would be enough for galaxies in nearby clusters (e.g. Virgo and Fornax) from either AGN (ideally) or ensembles of bright X-ray binaries.

\subsection{Intracluster Medium Microphysics} 

The ICM is a weakly collisional, highly magnetized plasma whose properties rule the thermodynamic evolution of galaxy clusters. Highly localized transport processes determine how the energy from AGN-driven feedback and hierarchical mergers is dissipated and thermalized across cosmic time \citep{Fabian2012, McNamara2012}. At present, we are lacking strong observational constraints on these crucial processes, introducing degeneracies and biases in cosmological hydrodynamical simulations; see e.g., BAHAMAS \citep{McCarthy2017},  Rhapsody-C \citep{Pellissier2023}, and FLAMINGO \citep{Schaye2023}. 
Ultra-high angular resolution X-ray imaging is the key to finally resolve these bottlenecks, as it will permit to measure spatially resolved diagnostics of the ICM.
Ultimately, resolving these small-scale plasma interfaces is the missing link required to properly use galaxy clusters for high-precision cosmology.

\subsubsection{Contact discontinuities}
\label{contactdis}
Despite decades of study with Chandra and other major X-ray missions, the microphysics of the ICM remains, in many respects, poorly understood \citep{ZuHone2016}.
With Chandra, ``contact discontinuities'' are already known to be sharper and smoother than simple hydrodynamic expectations, but current observations generally provide only a small number of resolution elements across the relevant structures \citep{Vikhlinin2001,Werner2016,ZuHone2016}.
The intrinsic widths, sharpness, and smoothness of these boundaries encode the suppression of thermal conduction and particle diffusion, the effective viscosity, and the role of magnetic draping \citep{Vikhlinin2001,MarkevitchVikhlinin2007,ZuHone2016,WangMarkevitch2018}.
These transport processes determine how long entropy, temperature, metallicity, and density substructure survive in the ICM, affecting the thermodynamic profiles and intrinsic scatter of cluster observables used in cosmological analyses.
For low-redshift clusters like Virgo and Perseus, a 50-200 mas uXRI would resolve the narrow interfaces where the weakly collisional ICM reveals its effective transport properties.
Ultra-high-resolution imaging would allow front widths and Kelvin–Helmholtz cutoff scales to be measured locally along many independent interface segments, converting qualitative evidence for suppressed transport into spatially resolved constraints on ICM microphysics.


\subsubsection{Shock fronts}
\label{shocks}

Merger shocks and AGN-driven weak shocks provide direct laboratories for understanding how energy is thermalized in the ICM. In mergers, shocks convert gravitationally driven bulk motions into heat, turbulence, and nonthermal particle populations \citep{MarkevitchVikhlinin2007,Russell2022,vanWeeren2019}.
In cool cores, repeated weak shocks and sound waves generated by AGN outbursts may distribute jet power through the cluster atmosphere \citep{Fabian2003,Fabian2006,Forman2007,Randall2011,Randall2015}.
Chandra observations of nearby cool cores such as Perseus and M87 reveal weak shocks, ripples, cavities, rims, and filaments on $\sim$kpc and sub-kpc scales \citep{Fabian2003,Fabian2006,Forman2007}, while deep observations of merger shocks such as Abell 2146 measure projected shock widths of order $\sim$10–20 kpc across several-hundred-kpc fronts \citep{Russell2022}. 
Resolving shocks together with cavities, rims, filaments, and contact discontinuities with a uXRI at 100 mas would directly measure how jet energy couples to the surrounding ICM \citep{Fabian2003,Birzan2004,McNamara2007,Randall2015}.
Such measurements would provide the microphysical closure needed to model how feedback and mergers set cluster pressure profiles, thermal histories, and the level of non-thermal support in cosmological simulations.
These topics directly relate to cosmology, because merger-driven turbulence, bulk motions, and non-thermal pressure support are major sources of scatter and bias in cluster mass estimates and in the mapping between dark-matter halo growth and observable ICM properties.

\subsection{nHz Gravitational Waves}

Many pulsars used in pulsar timing arrays \cite{2025Ap&SS.370..124T}
are also X-ray sources within 300 pc of Earth.
A uXRI with $\mu$as spatial resolution can measure accurate parallaxes to measure distances to a fraction of a gravitational wavelength.
With distance estimates to 0.1 pc, the correlations between the pulsar timing residuals can then be used to localize individual sources of nHz gravitational waves, rather than just the stochastic background\cite{2011MNRAS.414.3251L}.

\clearpage
\bibliographystyle{aasjournalv7}
\bibliography{references,referencesWG5,referencesWG3,referencesWG1and2}

\end{document}